\documentclass[aps,prl,reprint,superscriptaddress]{revtex4-2}
\usepackage{slashed}
\usepackage{amsmath}
\usepackage{amssymb}
\usepackage{graphicx}
\usepackage{mathtools}
\usepackage[caption=false,position=top]{subfig}
\usepackage[export]{adjustbox}
\usepackage{dsfont}
\usepackage{xcolor}
\usepackage{comment}
\usepackage{lineno}
\usepackage{orcidlink}
\usepackage[percent]{overpic}
\usepackage{hyperref}
\hypersetup{colorlinks=true,linkcolor=blue,citecolor=blue,urlcolor=blue}

\definecolor{darkgreen}{RGB}{0,100,0}

\usepackage[capitalize]{cleveref}

\def\be{\begin{equation}}
\def\ee{\end{equation}}
\def\bea{\begin{eqnarray}}
\def\eea{\end{eqnarray}}

\newcommand{\op}[1]{\hat{#1}}
\newcommand{\ck}{\op{c}^\dagger}
\newcommand{\ca}{\op{c}^{\phantom{\dagger}}}
\newcommand{\fk}{\op{f}^\dagger}
\newcommand{\fa}{\op{f}^{\phantom{\dagger}}}

\newcommand{\psik}{\op{\psi}^\dagger}
\newcommand{\psia}{\op{\psi}^{\phantom{\dagger}}}

\newcommand{\s}{\sigma}

\newcommand{\e}{\varepsilon}
\newcommand{\Spin}{\mathbf{\mathcal{S}}}

\newcommand{\vk}{\mathbf{k}}
\newcommand{\vq}{\mathbf{q}}
\newcommand{\pd}{\phantom{\dagger}}
\newcommand{\hc}{\mathrm{H.c.}}

\newcommand{\logicsec}[1]{\textcolor{black}{\emph{#1}}.---}

\begin{document}

\title{Kondo singlet from ferromagnetic coupling:\ an analog of Anderson-Morel superconductivity in the magnetic channel}

\author{Ewan Scott\,\orcidlink{0009-0002-8452-2262}}
\email{ewan.scott.18@ucl.ac.uk}
\affiliation{\hbox{Department of Mathematics, University College London, Gordon St., London WC1H 0AY, United Kingdom}}

\author{Yaqi Chen\,\orcidlink{0009-0007-2003-6501}}
\affiliation{Department of Mathematics, King’s College London, Strand, London WC2R 2LS, United Kingdom}

\author{Michael Turaev\,\orcidlink{0009-0003-8566-5078}}
\email{mturaev@uni-bonn.de}
\affiliation{Physikalisches Institut, Universit\"{a}t Bonn, Nussallee 12, 53115 Bonn, Germany}

\author{Tarkan Yzeiri\,\orcidlink{0009-0006-7831-6154}}
\affiliation{\hbox{London Centre for Nanotechnology, University College London, Gordon St., London, WC1H 0AH, United Kingdom}}
\affiliation{\hbox{ISIS Facility, Rutherford Appleton Laboratory, Chilton, Didcot, Oxfordshire OX11 0QX, United Kingdom}}

\author{Chris Hooley\,\orcidlink{0000-0002-9976-2405}}
\affiliation{Centre for Fluid and Complex Systems, Coventry University, Coventry CV1 2TT, United Kingdom}

\author{Krzysztof P. W{\'o}jcik\,\orcidlink{0000-0002-8201-1824}}
\email{kpwojcik@ifmpan.poznan.pl}
\affiliation{Institute of Molecular Physics, Polish Academy of Sciences, 
			 Smoluchowskiego 17, 60-179 Pozna{\'n}, Poland}

\author{Micha\l{} P. Kwasigroch\,\orcidlink{0000-0002-6613-2183}}
\affiliation{\hbox{Department of Mathematics, University College London, Gordon St., London WC1H 0AY, United Kingdom}}
\affiliation{Trinity College, Cambridge, CB2 1TQ, United Kingdom}

\date{\today}

\begin{abstract}
We consider magnetic impurities coupled to a conduction sea via a fully isotropic ferromagnetic spin-exchange term, the strength of which depends on the conduction-electron modes involved in the scattering. In the single-impurity case we show both analytically and numerically that there exists a parameter regime in which the conventional Kondo effect develops at low temperatures, leading to a singlet ground state. In the case of a lattice of impurities, we show that this leads to a heavy Fermi liquid state that is energetically favored over magnetic ordering in a broad parameter range. We argue that these effects are analogs of Anderson-Morel superconductivity, and discuss routes to their experimental realization.
\end{abstract}

\maketitle

\logicsec{Introduction}%
An important class of problems in physics concerns the coupling between localized degrees of freedom and delocalized ones. A paradigmatic example is the Kondo problem \cite{hewson,Coleman_2015}, the simplest version of which consists of a localized spin-$\frac{1}{2}$ particle (sometimes called an impurity) coupled via spin exchange to a continuum of delocalized conduction electrons.

In the case where this exchange interaction is antiferromagnetic, it was realized many decades ago that this problem exhibits a simple version of asymptotic freedom \cite{Anderson_1970}: the effective coupling between the impurity and the conduction sea \emph{increases} as the energy scale is lowered. This phenomenon, known as the Kondo effect, motivated intensive theoretical work in the 1960s and 1970s \cite{kondo,PhysicsPhysiqueFizika.2.5}, the highlight of which was perhaps the development of the numerical renormalization group (NRG) \cite{WilsonNRG}. The upshot of this analysis was that the growth in coupling is associated with the formation of coherent singlet correlations between the impurity and the conduction electrons, culminating at low energies in the complete screening of the impurity's magnetic moment and the recovery of Fermi-liquid behavior \cite{nozieres74}.

It was already clear from early renormalization group (RG) treatments of the problem that whether a Kondo singlet develops depends on the sign of the magnetic exchange coupling, $J$, between the impurity and the conduction sea \cite{andersonyuval69,yuvalanderson70,ayh70}. If the coupling is antiferromagnetic (AFM, positive), then the model flows to strong coupling and a singlet forms; if it is ferromagnetic (FM, negative), then it does not. This statement is of course an over-simplification:\ for example, if one relaxes the SU(2) symmetry of the coupling between the impurity and the conduction sea, one finds that a model in which all components of $J$ are negative can also renormalize to strong coupling \cite{Anderson_1970}, and more exotic fixed points are present in multi-channel scenarios \cite{Cox1998Sep}.

What, then, is the actual sign of $J$? In the simplest models, such as the Anderson impurity model, it can be shown to be antiferromagnetic. However, in more realistic models of materials there are several routes to ferromagnetic exchange between the impurity and the conduction sea:\ strong Hund's interaction \cite{Yin2011Dec,Georges2013Apr,Walter2020Sep,Hazra2023Mar,Dani2025May}; residual coupling between the partially screened impurity moment and the conduction sea in underscreened Kondo systems \cite{Nozieres1980Mar,Parcollet1997Dec,Varma2002May,Perkins2007Sep,Borda2009Mar,Girovsky2017May}; 
and various types of superexchange interaction in molecular systems and elsewhere \cite{Anderson1950Jul,Hackl2009May,Baruselli2013Jul,Turco2026Apr}.
Despite that, the FM case seems to be rarely discussed, and widely believed to be completely 
irrelevant for cases where strong-coupling features are apparent. With the present Letter we challenge this assessment.

From the late 1970s onwards, consideration of the Kondo effect was extended to include systems where the localized spins form a regular network, so-called Kondo lattice models \cite{DONIACH1977231,Stewart2001Oct}. These were motivated by physical settings such as the heavy fermion intermetallics, where rare-earth ions host localized moments that interact with the itinerant conduction electrons of the host metal. These materials show quantum phase transitions between magnetically ordered and heavy Fermi liquid states that are often interpreted in terms of a competition between Kondo screening of the individual moments and RKKY interactions between them \cite{Paschen2021Jan}. The Kondo-RKKY criticality can support various superconducting phases of notoriously intriguing properties 
\cite{Tachiki1984Mar,Coleman1988Dec,Bodensiek2013Apr,Nica2022Oct,Levitin2025Feb}.
As in the case of the single impurity, a microscopic ferromagnetic exchange is usually not considered as a possible mechanism behind the emergence of the heavy Fermi liquid phase in a Kondo lattice model.
In this Letter, we take a closer look at this possibility.

There is an analogy between the formation of the Kondo singlet in the Kondo problem and the formation of a condensate of Cooper pairs in superconductors. In this analogy, an antiferromagnetic interaction between the impurity spin and the conduction sea in the Kondo case corresponds to an attractive interaction between the electrons in the superconducting case. It is known, however, that attractive interactions are not a necessary condition for superconducting order to develop:\ as shown by Morel and Anderson \cite{AM}, a dependence of the repulsive interaction strength on $\vk$, the wavevector of the conduction electron mode, is enough to provoke a superconducting instability.
Motivated by this observation, we present in this Letter a single-impurity Kondo model in which the analogous phenomenon occurs. The model is fully SU(2)-symmetric and has only negative (FM) spin-exchange couplings; nonetheless, there are regions of parameter space in which a Kondo singlet forms in the ground state.

\begin{figure}[t!]
    \centering
    \begin{overpic}[width=\linewidth]{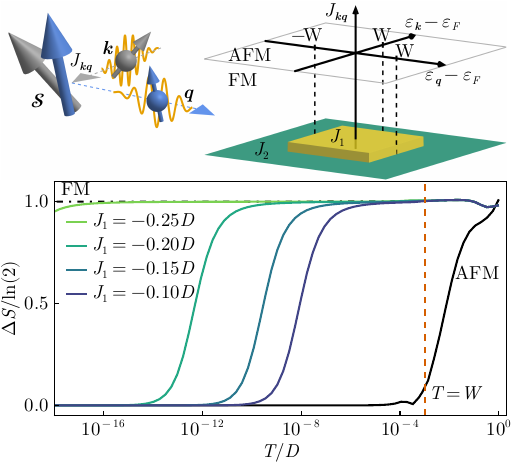}
    \put(0,85){(a)}
    \put(35,85){(b)}
    \put(0,50){(c)}
    \end{overpic}
    \caption{(a) Schematic illustration of an electron with momentum $\vk$ 
             scattering off a localized spin $\mathcal{S}$ to momentum spin $\vq$.
             (b) A plot of the assumed structure factor $J_{\mathbf{ kq}}^{\phantom{\dagger}}$
             as a function of Bloch-state energies $\e_{\vk}$ and $\e_{\vq}$.
             (c) Impurity contribution to entropy calculated from NRG for $W=10^{-3}D$, 
             $J_2=-0.5D$ and different $J_1$. 
             Lines denoted with AFM/FM correspond to $J_1=J_2=\pm0.5D$. 
             }
    \label{fig:schematic}
\end{figure}

\logicsec{Model}%
The Kondo coupling is conventionally derived from a local Coulomb repulsion via the Schrieffer-Wolff transformation \cite{Schrieffer1966Sep}. For a single-orbital impurity, the resulting exchange is AFM, with a momentum dependence that is usually disregarded. FM exchange generated via Hund's interactions or non-trivial orbital structure can likewise retain a momentum dependence. Here, we analyse a minimal model of a spin-$1/2$ impurity coupled to a host via momentum-dependent FM exchange. The Hamiltonian is
\begin{equation}
H_{\mathrm{K}} = \frac{1}{2}\sum_{\vk\vq\alpha\beta} J^{\pd}_{\vk\vq}  \op\Spin \cdot \mathbf{\sigma}^{\pd}_{\alpha\beta}  \ck_{\vq\alpha}\ca_{\vk\beta}
+\sum_{\vk\alpha}\e^{\pd}_{\vk}\ck_{\vk\alpha}\ca_{\vk\alpha}, 
\label{H1imp}
\end{equation}
where $J_{\vk\vq}$ is the exchange coupling, $\Spin$ is the impurity spin operator, $\mathbf{\sigma}$ is a vector of Pauli matrices, $\ck_{\vk\alpha}$ creates a conduction electron with momentum $\vk$ and spin $\alpha$, and $\e_{\vk}$ denotes the conduction electron dispersion. 

The key ingredient of the Anderson-Morel mechanism is that the interaction depends on the energies of the incoming and outgoing conduction electrons in a correlated manner. Motivated by the original potential~\cite{AM}, we take the exchange interaction to be
\begin{equation}
    J_{\vk\vq}= \begin{cases}
        J_1 & |\varepsilon_\vk -\varepsilon_F|,|\varepsilon_{\vq}-\varepsilon_F|<W \\
        J_2 & \mathrm{otherwise}
    \end{cases} \quad , \label{J12}
\end{equation}
with $\e_{\vk}$ giving a rectangular density of states. Here, $W < D$ defines a window around the Fermi level, $\varepsilon_F$, where the interaction strength is reduced but remains FM, i.e. $J_1, J_2<0$, with $|J_1| < |J_2|$, see \cref{fig:schematic}, where $D$ is the half-bandwidth of the conduction electrons. Crucially, this form is nonseparable, i.e. $J_{\vk\vq} \neq j_{\vk}j_{\vq}$, and therefore cannot be reduced to a single conduction-electron mode coupled to the impurity.

\begin{figure}[t!]
    \begin{overpic}[width=\linewidth]{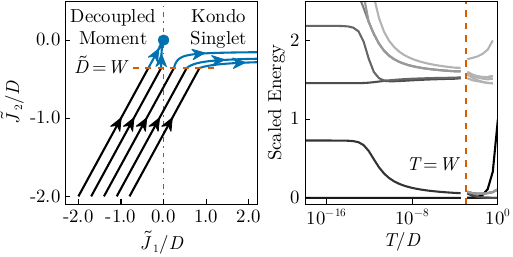}
    \put(2,47){(a)}
    \put(50.5,47){(b)}
    \end{overpic}
    \caption{(a) RG trajectories in the $J_1,J_2$ plane with $W   = 0.1D$. For visual guidance, the flow is schematically continued for $\tilde{D}<W$ (blue lines). (b) numerical RG flow (rescaled energies of many-body eigenstates at even iterations) for $J_1=-0.2D$, $J_2=-0.5D$, $W=0.001D$. }
    \label{fig:RG}
\end{figure}

\logicsec{Perturbative RG analysis}%
We first analyze the model using a standard perturbative RG procedure, successively integrating out high-energy states \cite{Anderson_1970}, thereby reducing the original cutoff $D$ to $\tilde{D}$. This renormalizes the exchange couplings, obeying
\begin{align}
\label{eq:PRG-eom}
    \tilde{D} \frac{d \tilde{J}_i}{d\tilde{D}} & = - \frac{\tilde{J}_2^2}{D}
\end{align}
for both $i=1,2$ (see End Matter for more details). \cref{fig:RG}(a) shows RG trajectories for different initial $J_1$. The scale $W$ naturally divides the flow into two regimes, (i) $D > \tilde{D} > W$ and (ii) $W>\tilde{D}$, in a similar spirit to the scaling analysis of the Anderson model \cite{Haldane}. For FM exchange without energy dependence, i.e. for $J_1=J_2$, the RG flow approaches the decoupled local moment fixed point, corresponding to $\tilde{J}_1 = \tilde{J}_2 = 0$ \cite{Anderson_1970}. In our case, however, if the difference of $\tilde{J}_1(D)$ and $\tilde{J}_2(D)$ is sufficiently large, $\tilde{J}_1$ can change its sign within the region (i) and become AFM. This indicates that AFM Kondo physics emerges, destabilizing the local moment fixed point. 

Once the cutoff is reduced below $\tilde{D}=W$, the distinction between high energy processes with $|\e_{\vk}| > W$ and low energy processes with $|\e_{\vk}| < W$ is no longer meaningful. Thus, \cref{eq:PRG-eom} no longer holds below this scale, and we terminate the RG flow at $\tilde{D}=W$. The formation of a Kondo singlet ground state is determined by whether the condition $\tilde{J}_1(W) > 0$ is reached in region (i), as this is the only mechanism that can drive the system toward the strong coupling fixed point. Solving \cref{eq:PRG-eom}, the requirement $\tilde{J}_1 > 0$ can be expressed as a condition on the bare couplings,
\begin{equation}
    J_1 > J_1^* = J_2 \left[ 1 - \frac{1}{1+(J_2/D)\ln(W/D)}\right].
    \label{eq:J1star}
\end{equation}
As in the Anderson-Morel model, the narrower the energy window $W$, the smaller the difference $|J_1-J_2|$ can be. However, this comes at the cost of a reduced Kondo scale, with the Kondo temperature $T_K$ analyzed after \cref{TK}.

\logicsec{Numerical RG results}%
The perturbative RG analysis provides analytic expressions and captures the RG flow in the weak-coupling regime. Once the strong-coupling regime is approached, the nature of the ground state can only be extrapolated. This motivates an NRG treatment of \cref{H1imp} with exchange interaction defined in \cref{J12}. The nonseparable form of $J_{\mathbf{kq}} \neq j_{\mathbf{ k}}j_{\vq}$ cannot be represented by a single local field operator per spin, and therefore requires a formulation beyond the standard single-channel NRG \cite{NRG_RMP} or its generalization to arbitrary conduction band spectra \cite{dosNRG,Zalom2023Nov}. Instead, the couplings $J_1$ and $J_2$ define two different field modes, of different scale of localization, both coupled to the impurity. A convenient orthogonal basis choice is $A_1\psia_{1\s} = \sum_{\vk: |\e_{\vk}|<W} \ca_{\vk\s}$ and $A_2\psia_{2\s} = \sum_{\vk: |\e_{\vk}|>W} \ca_{\vk\s}$, with positive normalization constants $A_i$. The exchange term can then be written as 
\begin{align}
    K  &= \sum_{\alpha\beta} \vec\Spin \cdot \vec\sigma_{\alpha\beta}  \bigg[ 
        \frac{J_1}{2} |A_1^2| \psik_{1\alpha}\psia_{1\beta} 
        + \frac{J_2}{2} |A_2^2| \psik_{2\alpha}\psia_{2\beta}  
        \nonumber\\& \quad 
        + \frac{J_2}{2} A_1A_2 ( \psik_{1\alpha}\psia_{2\beta} + \hc)\bigg].
        \label{K} 
\end{align}
An additional difficulty arises because the bandwidths of these channels have different orders of magnitudes, and the mode corresponding to $\psia_{2\s}$ has a gap, thereby spoiling the scale separation characteristic of NRG. This problem can be circumvented by first solving the system with a large band-width. The two-mode NRG formulation and the additional computational costs are discussed in the End Matter. As a benchmark, this procedure was tested for $J_1=J_2=J$, where the conventional single-channel result is recovered for different $W$ and both signs of $J$.

The central NRG result we obtain is that purely ferromagnetic bare exchange couplings can produce a spin-singlet ground state over an extended parameter regime. This can be seen directly from the impurity contribution to entropy in \cref{fig:schematic}(c), and in the fixed point spectrum in \cref{fig:RG}(b), shown for representative parameters. At high temperatures, the RG flow is very similar to the $J_1=J_2$ case. However, below the scale $W$, it deviates and flows toward the strong-coupling spectrum at the scale corresponding to $T_K$. Motivated by the perturbative RG results, we estimate the Kondo temperature by standard Kondo form taken at $\tilde{D} = W$. This gives
\begin{equation}
 T_K \approx W \exp\left[-\frac{D}{\alpha(J^{\phantom{*}}_1 - J_1^{*})}\right] ,
 \label{TK}
\end{equation}
where $J_1^{*}$ marks the threshold of the bare coupling $J_1$ such that the singlet is the eventual ground state for $J_1>J_1^*$, and both $\alpha$ and $J_1^{*}$ are fitting parameters. This form gives very good agreement over a wide parameter range, see End Matter for more details.

\begin{figure}
    \centering
    \begin{overpic}[width=\linewidth]{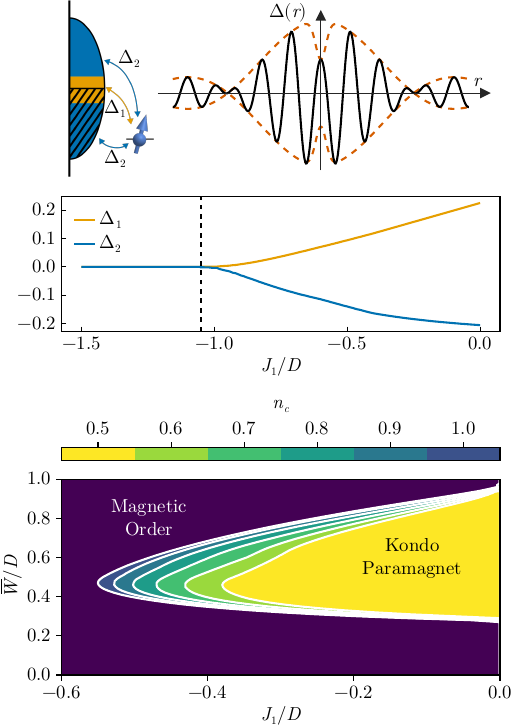}
        \put(1,98){(a)}
        \put(25,98){(b)}
        \put(1,73){(c)}
        \put(1,45){(d)}
    \end{overpic}
    \caption{Mean-field solution for the lattice case: 
            (a) A schematic illustration of the meaning of the two fields $\Delta_1$ and $\Delta_2$ that describe the hybridization between the local moments and the conduction sea in the mean-field approximation.
            (b) Schematic $\Delta(r)$ for a particular choice of $\Delta_1$ and $\Delta_2$ --- the analog of the gap function in the Anderson-Morel model.
            (c) $\Delta_1$ and $\Delta_2$ as functions of $J_1$ for $J_2=-2D$, $n_c=0.9$, $W=0.5D$. The vertical dashed line indicates the onset of a non-zero solution.  
            (d) competition between magnetic order and Kondo screening in the lattice, with $J_2 = -2D$. 
            }
    \label{fig:lattice}
\end{figure}
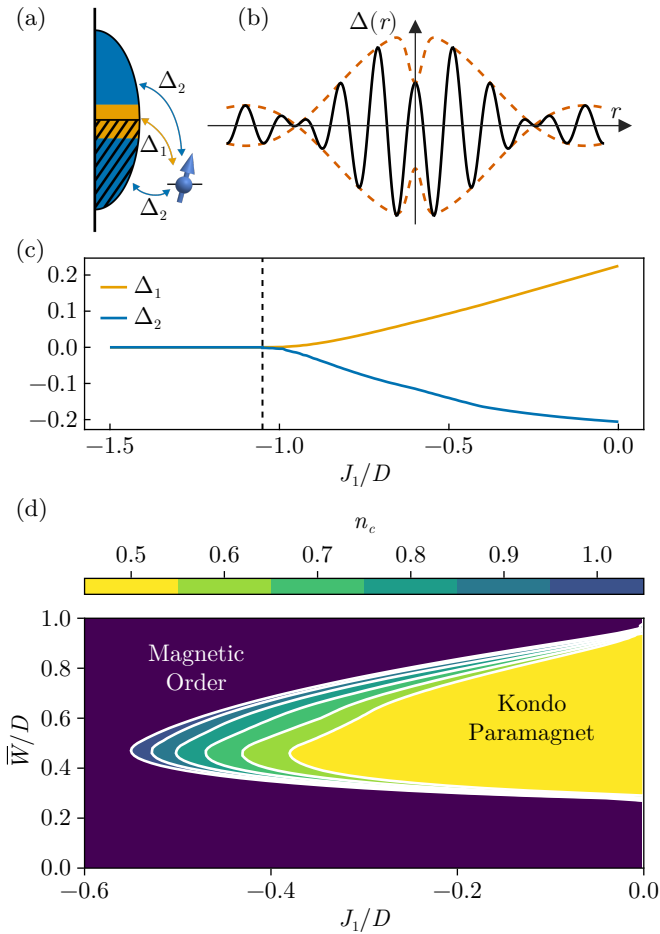

\logicsec{Lattice model and mean-field analysis}%
While the single impurity case can be solved numerically in a reliable manner, the lattice case is much more difficult. 
Therefore, in this section we utilize mean field theory to understand how the present coupling scheme, in close analogy with the Anderson-Morel mean field treatment of superconductivity, affects the Kondo lattice result.

To this end, we define the Kondo lattice Hamiltonian with FM couplings after fermionisation as \cite{Coleman_2015}
\begin{align} \nonumber
H_{\rm KL}=
    -\sum_{\vk\vq\alpha\beta}
    \frac{J_{\vk\vq}}{2}\left(\ck_{\vq\alpha}\fa_{\vq\alpha}\right)\cdot& \left(\fk_{\vk\beta}\ca_{\vk\beta}\right)
    +\sum_{\vk\alpha}\e_{\vk}\ck_{\vk\alpha}\ca_{\vk\alpha}
    \\ &
    -\sum_{\vk\alpha}\mu \fk_{\vk\alpha}\fa_{\vk\alpha}.
\end{align}
where $\fa_{\vk\alpha}$ ($\ca_{\vk\alpha}$) annihilates an auxiliary impurity (conduction) fermion with momentum $\vk$ and spin $\alpha$, $J_{\vk\vq}$ is given by \cref{J12} and $\e_{\vk}$ is the conduction band dispersion 
and includes a shift to fix the density of conduction electrons, $n_c$. The Lagrange multiplier, $\mu$, is introduced to fix the constraint on the occupancy of the impurity fermions, $\sum_{\vk\alpha} \fk_{\vk\alpha}\fa_{\vk\alpha}=N$, where $N$ is the number of sites. We then introduce the Kondo hybridization order parameter and derive the mean field Hamiltonian
\begin{align}\label{MFHamiltonian}
H_{\rm MF}&
    =\sum_{\vk\alpha}\left[\e_{\vk}\ck_{\vk\alpha}\ca_{\vk\alpha}
    -\mu \fk_{\vk\alpha}\fa_{\vk\alpha}\right]+
    \\& \nonumber\quad\quad\quad\quad\quad
    -\sum_{\vk\alpha}\left( \Delta_{\vk}\fk_{\vk\alpha}\ca_{\vk\alpha} 
    +\hc \right), \\
    \label{DeltaEQ}
\Delta_{\vk}&=
    \frac{1}{2}\sum_{\vq\alpha} J_{\vk\vq}
    \left\langle \ck_{\vq\alpha}\fa_{\vq\alpha}\right\rangle.
\end{align}
The hybridization field $\Delta_{\vk}$ is the mean-field order parameter; we find mean-field solutions in which $\Delta_{\vk} = \Delta_1$ when $|\e_{\vk}|<W$ and $\Delta_2$ otherwise.  This is represented schematically in Fig.~\ref{fig:lattice}(a).

In Fig.~\ref{fig:lattice}(b) we show the resulting form of the hybridization in real space, $\Delta(\mathbf{r})$, in the continuum limit and with $r=|\mathbf{r}|$. The strong analogy with Anderson-Morel superconductivity is apparent. In the superconducting case, Coulomb repulsion creates a sharp dip in the gap function $\Delta_{\rm gap}(\mathbf{r})$ at $\mathbf{r}=\mathbf{0}$. The Cooper pair wavefunction is modified as a result to lower the local Coulomb energy penalty, allowing the BCS state to form. 
 Analogously, we have shown that strong ferromagnetic exchange (e.g.\ from a strong local Hund's coupling) can be overcome, and a Kondo phase formed, when the hybridization order parameter $\Delta(\mathbf{r})$ acquires a sharp dip at $\mathbf{r}=\mathbf{0}$. 
 Fig.~\ref{fig:lattice}(c) shows the solution to the mean field self-consistency equations for $\Delta_1$ and $\Delta_2$ as a function of $J_1$. As $J_1$ increases a nonzero solution becomes stable, indicating the existence of a heavy-fermion groundstate.
 
 To verify the Kondo phase as the groundstate of the system, we compare the energies of the Kondo and FM ordered phases and find that the Kondo phase wins over a large range of parameters. Our results are presented in Fig.~\ref{fig:lattice}(d) for a range of conduction electron fillings $n_c$, and various choices of the model parameters defined in \cref{fig:schematic}(b).
 Since the Fermi surface moves as the conduction electron filling is varied, we have introduced the parameter $\bar{W}/D$, equal to the proportion of the total bandwidth that $J_1$ acts on, see End Matter for details.

For a standard AFM coupling between the impurity and conduction electrons, the competition between RKKY exchange and coherent Kondo screening leads to the well established Doniach phase diagram. The magnetically ordered phase is preferred at weak dimensionless coupling and the Kondo paramagnet once it becomes strong enough. In our model, with purely FM coupling, an unusual inversion of the standard Doniach phase diagram takes place; the  Kondo paramagnetic phase is now preferred when the dimensionless coupling at the Fermi level ($J_1/D$) is weak enough, rather than strong enough.

\logicsec{Discussion}%
In this Letter we have demonstrated that a single impurity coupled to a conduction sea by purely ferromagnetic couplings can nonetheless become screened via the Kondo effect in the widely applicable case of energy-dependent couplings. This is a direct analog of  Anderson-Morel superconductivity, in which Cooper pair formation occurs via an energy-modulated but always repulsive interaction between the electrons.

How should we understand this surprising result?  It stems from the fact that, while the ferromagnetic coupling decays logarithmically with decreasing energy scale and is thus irrelevant in the infrared limit, it therefore grows logarithmically with increasing energy scale and becomes relevant in the ultraviolet limit. This allows a ferromagnetic coupling at higher energies to strongly renormalize the ferromagnetic coupling at the Fermi level, and even invert its sign. Our result points to the importance of considering higher-energy processes in the presence of ferromagnetic couplings at the Fermi level.

We have also shown that a similar phenomenon occurs for a Kondo lattice. In this setting it corresponds to the impurities virtually enlarging the Fermi surface, and increasing the Fermi volume in accordance with the Luttinger sum rule.  We have shown that, although the energy of the resulting Kondo paramagnet is significantly higher than in the case of purely antiferromagnetic Kondo couplings, it can still outcompete the magnetically ordered state in a large region of parameter space, establishing an analog of Doniach phase diagram for ferromagnetic couplings.

We believe that more intricate screening scenarios could be realized along the lines presented in this Letter:\ these would be analogous to more 
sophisticated models of superconductivity, with potential applications to the heavy-fermion compounds.
Our results may also have implications for underscreened Kondo models, 
where the residual spin is coupled ferromagnetically to the conduction band. 
Our findings indicate that, contrary to common belief, with appropriate energy structure 
of this coupling a further screening at even lower energy scale cannot be excluded, even
in a complete absence of additional screening channels. 

There remains the question of whether the effect predicted in this Letter can be observed experimentally. One intriguing possibility is that it already has been:\ it is possible that cases like this one have been misdiagnosed as impurities with antiferromagnetic couplings on the grounds that Kondo screening was seen to occur. 
It may also be considered a candidate explanation in cases where Kondo features are seen unexpectedly \cite{Fuhrman2018Apr,Li2020Sep,Pirie2023Mar}.
We hope that our observations in this Letter will stimulate work, both theoretical and experimental, to ascertain the sign of the microscopic coupling between the conduction sea and the impurity directly.

\begin{acknowledgments}
\logicsec{Acknowledgments}%
Stimulating discussions with 
Piers Coleman,
Gilbert Lonzarich,
and Jan Martinek
are gratefully acknowledged. M.T. acknowledges funding by the 
Deutsche Forschungsgemeinschaft Collaborative Research Center CRC 185 OSCAR (No. 277625399). 
KPW was supported by National Science Centre in Poland 
through grant no.~2023/51/D/ST3/00532
and acknowledges computing time at Pozna\'n Supercomputing and Networking Center.

\logicsec{Data availability}%
The data that support the findings of this article are openly available  \cite{ZenodoData}.
\end{acknowledgments}

\bibliography{references}

\onecolumngrid
\pagebreak[4]
\begin{center}
{\huge End matter}
\end{center}
\twocolumngrid

\logicsec{Perturbative RG details}%
In the considered model, there are in principle $3$ couplings undergoing the 
flow for $D>\tilde{D}>W$: low-energy exchange $J_1$, high-energy exchange $J_2$,
and the term $J_3$ corresponding to an incoming low-energy electron with $|\varepsilon_{k}| < W$ 
and outgoing high-energy electron with $|\varepsilon_{k}| > W$ (or vice versa).
However, since $J_3$ and $J_2$ share identical initial conditions, it can be 
shown that they remain equal throughout the RG flow. Therefore, the coupling 
$J_3$ can be trivially eliminated, resulting in \cref{eq:PRG-eom}.

\begin{figure}[tb!]
    \centering
    \begin{overpic}[width=\linewidth]{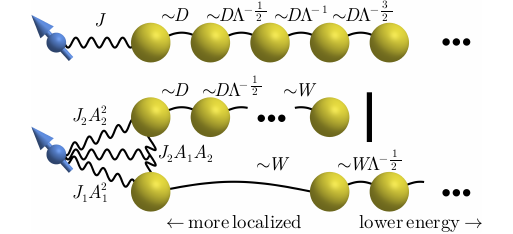}
    \put(0,40){(a)}
    \put(0,25){(b)}
    \end{overpic}
    \caption{Illustration of effective NRG Hamiltonians:
        (a) conventional Wilson chain for Kondo model with $J_{\vk\vq} = J$ .
        (b) Wilson chain corresponding to exchange interaction given by \cref{K}.
        Yellow balls indicate non-interacting sites,
        small blue ball -- an impurity,
        wavy lines the Kondo interaction terms,
        and solid lines simple hopping terms.}
    \label{fig:WilsonChains}
\end{figure}

\logicsec{Numerical RG details}%
The basis for NRG discretization is \cref{K}. It allows to rewrite \cref{H1imp} 
in a form identical to a $2$-channel Kondo model with an inter-channel coupling,
\begin{align}
H_K = &
    \sum_c J_c \, \op{\vec\Spin}\,\cdot \op{\vec{s}}_c
        + \sum_{c\s} \sum_{\vk\in\{\vk\}_c} \e_{\vk} \ck_{\vk\s}\ca_{\vk\s}
        \nonumber\\& \quad 
        + \sum_{\alpha\beta} \frac{J_{AB}}{2} \op{\vec\Spin}\cdot( \psik_{A\alpha}\vec{\s}_{\alpha\beta}\psia_{B\beta} + \hc) ,
\end{align}
where $c$ runs through channels $A$ and $B$, $J_A=J_1 A_1^2$, $J_B = J_2 A_2^2$, $J_{AB}=J_2 A_1A_2$,
and $\{\vk\}_A$ ($\{\vk\}_B$) is a subset of momenta for which $|\e_{\vk}|<W$ ($|\e_{\vk}| \geq W$). 
Since these two sets are disjoint and the kinetic energy term is diagonal in momenta, 
we can proceed with a standard NRG discretization scheme in each of them 
by going to continuum limit and identifying the densities of conduction band states 
$\rho_c(\omega) = \sum_{\vk\in\{\vk\}_c} \delta(\omega-\e_{\vk})$ \cite{NRG_RMP,dosNRG}. 
In fact, it is easy to observe that $\rho_A(\omega) = \rho(\omega) \theta(W-|\omega|)$
and $\rho_B(\omega) = \rho(\omega) \theta(|\omega|-W)$ with the Heaviside step function $\theta$. This means that channel $B$ has 
the original bandwidth and a gap of the size of $W$, while $W$ serves also as a bandwidth 
for gapless channel~$A$. 

Standard NRG discretization maps the original Kondo Hamiltonian onto the tight-binding 
$1$-dimensional half-infinite model, so-called Wilson chain, where the first site is 
the original impurity. The second site corresponds to a local conduction electron field 
at the impurity. The following sites are coupled with hoppings and on-site 
energies decaying exponentially along the chain (the first hopping is of the order of $D$),
and constitute an effective bath. This is illustrated in \cref{fig:WilsonChains}(a).
The rate of decay is determined by the discretization parameter $\Lambda$, 
with $2 \leq \Lambda \leq 3$ frequently used in practical calculations and $\Lambda \to 1$ 
corresponding to the continuum limit.

In the present case, a discretization of the gapless channel is straightforward and 
for rectangular density of states of the original band leads to Wilson chain identical
to the regular NRG case, just with all hoppings and on-site energies rescaled by 
factor $W/D$. 
A little more care has to be given to the discretization of a gapped channel $B$. 
Typically gapped hosts are somewhat problematic, because the original Wilsonian 
logarithmic discretization gives only a finite chain, whose smallest coupling 
corresponds to the gap itself. To come closer to the continuum limit, it is then 
advisable that the logarithmic refinement of discretization is shifted to the gap 
edge. Then, extra care has to be taken to realize iterative scale-by-scale solution
of the model \cite{Zalom2023Nov}. The benefit is that the spectral features around 
the gap edge are more precisely resolved, which is crucial for accurate calculation 
of low-temperature properties in the absence of low-energy excitations. 
Here, however, we do not need to boost the accuracy near the band edge, because channel $A$
does, indeed, supplement low-energy excitations and is certain to dominate low-temperature
physics. In turn, as explained in the following, including too many states to describe
high-energy part of the Hamiltonian could in fact deplete computational possibilities.
Thus, we keep the original Wilsonian discretization and end up with an effective model 
whose schematic is depicted in \cref{fig:WilsonChains}(b), with channel $A$ shifted to
the right to indicate its diminished energy scale.

It should be noticed that to keep the scale-by-scale solution approach, the defining 
feature of the RG procedures, one has to solve the finite channel $B$ first, and couple 
the low-energy channel $A$ later on. This, however, causes another difficulty, namely 
that the first hopping in channel $A$ couples low-energy part of the Hamiltonian directly 
to the (high-energy) direct vicinity of the impurity. 
Therefore, one has to not only calculate matrix elements of $\psi_{A\s}$ between the states
kept till the last iteration of diagonalization of channel $B$, but also take care to keep 
the states whose matrix elements are indeed large through all procedure of diagonalization of channel $B$, even if their excitation energy is sizeable.
Our solution, inspired by DMRG ideas \cite{Schollwock2005Apr}, is to modify the criterion 
for keeping states while performing iterative solution of chain $B$ such, that additionally 
to the states that would be kept in traditional NRG, $|k_1\rangle, \ldots, |k_{N}\rangle$, we also 
keep an equal number of would-be-discarded states, $|d_{i_1}\rangle, \ldots, |d_{i_N}\rangle$, 
with largest indicators 
\begin{equation}
I_{j_m} = \sum_{i} \left( |\langle d_{j_m} | \psia_{A\s} | k_i \rangle|^2 + |\langle d_{j_m} | \psik_{A\s} | k_i \rangle|^2 \right) ,
\end{equation}
as long as there are so many. On top of that, we additionally increase the number of kept
states to avoid splitting near-degenerate states (differing in rescaled energy by less 
then $dE=10^{-4}$) between kept and discarded states, also for these additionally kept states.
We find it important to explicitly verify that all pairs of states connected by $\Spin_z$
reflection are either both kept or both discarded. 
Optimization of this $2$-mode NRG procedure is an interesting direction of future numerical studies.

In the calculations presented here we use $\Lambda=3$, $N=1500$, and assume rectangular 
density of states for the original conduction band. We note that in the computation scheme
outlined above also the $J_1=J_2$ case leads to (then artificial) division of the 
conduction band into two channels. This allows for testing if this division does not 
lead to any artificial results. 
We directly verify that the $\Delta S(T)$ for $J_1=J_2=\pm 0.5$
obtained from the NRG presented here does not deviate from one obtained for standard NRG 
by more than a few percent. 
In particular, a fragile flow to the local moment fixed point is correctly recovered.
\begin{figure}[!tb]
    \centering
    \includegraphics[width=\linewidth]{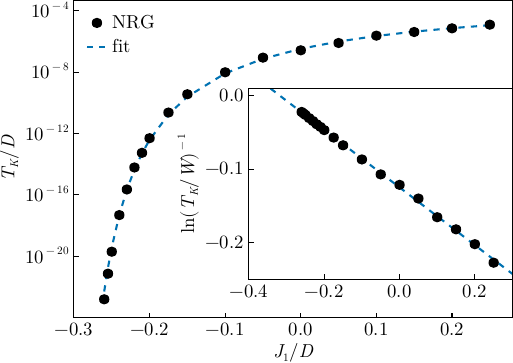}
    \caption{The Kondo temperature as a function of $J_1$ for $J_2 = -0.5D$ and $W=0.001$ calculated from NRG. The dashed line indicate the fit giving the values for $\alpha_j$ in \cref{TK}.}
    \label{fig:TK}
\end{figure}

\logicsec{The Kondo temperature}
The Kondo temperature has been obtained from entropy temperature dependence $\Delta S(T)$ through the requirement $\Delta S(T_K) = \ln(2)/2$. Then, the dependence of $\ln(T_K/W)$ on $J_1$ has been fitted with \cref{TK} by the least square method. The dependence of $T_K$ on $J_1$ is shown in \cref{fig:TK}, together with a fit. Note that both NRG results and the fit remain continuous through $J_1=0$. The critical $J_1$ obtained from the fit 
is $J_1^* \approx -0.313D$, whereas perturbative RG predicts $J_1^*\sim -0.388D$ through \cref{eq:J1star}. Existence of such a correction should not be surprising, given the fact that we use rather large bare coupling $J_2=-0.5D$, and the perturbative approach treats $J_2/D$ as a small parameter and is done in the leading order only. For the coefficient $\alpha$, we obtain $\alpha \approx 0.413$, while from conventional $1$-channel NRG for the Kondo 
model we have got $\alpha \approx 0.465$.
$\alpha \neq 1$ can be understood as a correction to the $1$-loop perturbative RG estimation of $T_K$. Note that this estimation is based on divergence in the flow that only appears in the perturbation theory, while we define $T_K$ through physically meaningful $\Delta S(T)$
dependence. The fact that $T_K$ follows \cref{TK} so well provides further argument for 
trusting the interpretation that for $\tilde{D}<W$ we see a standard Kondo RG flow characteristic of AFM coupling.

\logicsec{Lattice model and mean-field details}
After defining the Kondo hybridisation parameter from \cref{DeltaEQ}, one can derive the eigenvalues and eigenvectors:
\begin{eqnarray}\label{MFEig}
    E^{\pm}_{\vk} &=& \frac{(\e_{\vk} - \mu) \pm \sqrt{(\e_{\vk} + \mu)^2+4|\Delta_{\vk}|^2}}{2} \\ \nonumber u^{\pm}_{\vk}\! &=&\! \frac{-1}{\sqrt{1+|\Delta_\vk|^2/(E^{\pm}_{\vk}+\mu)^2}},\quad v^{\pm}_{\vk} = \frac{-\Delta_\vk u_{\vk}^\pm}{E^{\pm}_{\vk}+\mu}.
\end{eqnarray}
From which one can express the Kondo hybridization parameter in terms of the eigenvectors that diagonalise the Hamiltonian to obtain (for $n_{\mathrm{c}}\leq1$ at zero temperature):
\begin{equation}\label{GapEqn}
\Delta_{\vk}=\rho \int_{-D}^{D}J_{\vk\vq}\frac{\Delta_{\vq}}{\sqrt{(\e_{\vq}+\mu)^2+4|\Delta_{\vq}|^2}} \Theta(-E^-_{\vq})d\e_{\vq}
\end{equation}
where we have used a constant density of states $\rho = (2D)^{-1}$. And similarly for the density of impurity electrons, $ n_{\mathrm{f}} = \frac{1}{N}\sum_{\mathrm{i},\alpha} \langle \op{f}^\dagger_{i\alpha} \op{f}_{i\alpha}\rangle$, one can derive the following constraint:
\begin{equation}\label{nfMETAL}
 n_{\rm f} =\rho \int_{-D}^{D}\left[
1+\left(
1+\frac{4|\Delta_\vk|^2}{(\e_\vk+\mu)^2}
\right)^{-1/2}
\right] \Theta(-E^-_{\vq})d\e_{\vq}.
\end{equation}
\cref{GapEqn,nfMETAL} can then be used to self-consistently solve for $\Delta_1$, $\Delta_2$, and $\mu$. In order to compare results for a varying density of conduction electrons, $n_{\mathrm{c}}$, we now represent the width of the region with coupling $J_1$ using the variable $\bar{W}$:
\begin{equation}
    \bar{W} = \frac{1}{2}
    \int^{D}_{-D} \Theta\left( W - |\e_\vk - \e_{\mathrm{F}}| \right) d\e_\vk
\end{equation}
Where the Fermi energy of the unhybridised conduction electron bands, $\e_{\mathrm{F}}$, can be calculated by imposing:
\begin{equation}
    n_{\mathrm{c}} = \rho \sum_{\sigma}\int^{D}_{-D}\Theta\left(-E_{\sigma}(\e_\vk)\right) d\e_\vk,
\end{equation}
with $E_{\sigma} = \e_{\vk} + \lambda$, where $\lambda$ is a Lagrange multiplier introduced to fix the density of conduction electrons. The Fermi energy is then
\begin{equation}
    \e_{\mathrm{F}} = -\left(1-n_{\rm c} \right)D.
\end{equation}

\end{document}